\documentclass[aps,prl,twocolumn,groupedaddress,showpacs]{revtex4}

\usepackage{hyperref}
\usepackage{dcolumn}
\usepackage{xspace}
\usepackage{amssymb}
\usepackage{amsmath}
\usepackage{graphicx}
\usepackage{bm}

\def\tri{{{}^3{\rm H}}}

\def\het{{{}^3{\rm He}}}
\def\heq{{{}^4{\rm He}}}

\def\pham{\phantom{-}}
\def\phan{\phantom{9}}

\begin{document}

\title{Effect of three-nucleon interaction in $p-\het$ elastic scattering}

\author{M. Viviani$^1$,  L. Girlanda$^{2}$, A. Kievsky$^1$, and L.E. Marcucci$^{1,3}$}

\affiliation{
$^1$  Istituto Nazionale di Fisica Nucleare, Sezione di Pisa, 
Largo B. Pontecorvo 3, I-56127, Pisa, Italy \\ 
$^2$ Department of Mathematics and Physics, University of Salento, and
INFN-Lecce, I-73100 Lecce, Italy \\
$^3$ Department of Physics University of Pisa, 
Largo B. Pontecorvo 3, I-56127, Pisa, Italy }

\begin{abstract}
We present a detailed study of the effect of different three-nucleon
interaction models in $p-\het$ elastic scattering at low energies. 
In particular, two models have been considered: one derived from
effective field theory at next-to-next-to-leading order and one
derived from a more phenomenological point of view -- the 
so-called Illinois model. The four-nucleon scattering observables are 
calculated using the Kohn variational principle and the
hyperspherical harmonics technique and the results are compared with
available  experimental data. We have found that the
inclusion of either one of the other force model improves the
agreement with the experimental data, in particular for the proton
vector analyzing power. 
\end{abstract}

%

\maketitle



The complete knowledge of the three-nucleon (3N) interaction is 
one of the open question in Nuclear Physics nowadays. As is well known, there exist a
number of different realistic nucleon-nucleon (NN) interaction models
capable to reproduce almost perfectly the experimental NN scattering data up to
energies of 350 MeV. However, with only this
component of the nuclear interaction, one encounters several problems
in the description of $A\ge 3$ nuclear systems (see,
e.g., Refs.~\cite{Gea96,CS98,KEMN12}).  
To improve that situation, different 3N forces have been introduced.  

The recent development of 3N forces has followed
mainly two lines. First, there are 3N force models derived within a
chiral effective field theory (EFT) approach~\cite{Kolck94,Eea02}. Models
derived at next-to-next-to-leading order (N2LO) of the
so-called chiral expansion have been used so far. At this particular order, the 3N force
contains two unknown constants~\cite{Eea02} usually determined either by fitting
the 3N and four-nucleon (4N) binding energies~\cite{N07} or, alternatively, the 3N binding
energy and the Gamow-Teller matrix element (GTME) in the tritium
$\beta$-decay~\cite{Gazit,Mea12}. The 3N force depends also on a
cutoff function,   
which in general includes a cutoff parameter $\Lambda$. With a
particular choice of the cutoff function, a local version of the N2LO
3N interaction has been derived~\cite{N07}. The parameter $\Lambda$ is chosen to be for
physical reason of the order of 500 MeV (for a discussion about the 
size of the $\Lambda$, see Ref.\cite{ME12}). The derivation of chiral 3N
force at successive orders is now in rapid progress~\cite{BEKM11,KGE12,GKV11}.   

Alternatively, within a more phenomenological approach, the so-called
Illinois model for the 3N force model has been
derived~\cite{PPWC01}. This model has been constructed to include 
specific two- and three-pion exchange mechanisms between
the three nucleons. The model contains a few unknown parameters, which
have been determined by fitting the spectra of $A=4-12$ nuclei.  

Clearly, it is very important to test these models to understand 
how they describe nuclear dynamics. The $A=3$ and
$4$ scattering observables are between the best testing grounds to
this aim. However, most of the $A=3$ scattering observables are not
very sensitive to the effect of the 3N force~\cite{Gea96,KEMN12}. It is
therefore of relevance to study their effect in 4N systems.

In recent years, there has  been a rapid advance in solving the 4N
scattering problem with realistic Hamiltonians. Accurate
calculations of four-body scattering observables have been achieved in
the framework of the Alt-Grassberger-Sandhas (AGS)
equations~\cite{DF07,DF07b}, solved in momentum space, where the
long-range Coulomb interaction is treated using the
screening-renormalization me\-thod \cite{Alt78,DFS05}. 
Solutions of the Faddeev-Yakubovsky (FY) equations in configuration
space~\cite{Cie98,Lea05} and several calculations using the  
resonating group model \cite{HH03,Sofia08} were also reported.  In this
contribution, the four-body scattering problem is solved using the
Kohn variational method and expanding the internal part of the wave
function in terms of the hyperspherical harmonic (HH) functions (for a
review, see Ref. \cite{rep08}).   
Very recently, the efforts of the various groups have culminated 
in a benchmark paper~\cite{Vea11}, where it was shown that $p-\het$
and $n-\tri$  phase-shifts calculated using the AGS, FY, and HH
techniques and using several types of NN potentials are in very close
agreement with each other (at the level of or less than 1\%).   

Since 4N scattering observables can be calculated with high
accuracy, it is  timely to investigate the effect of
the 3N force in these systems. It is important to note that
the 4N studies performed so far have revealed the presence of several
discrepancies between theoretical predictions and experimental
data. In $p-\het$ elastic scattering several accurate
measurements exist for the unpolarized cross 
section~\cite{Fam54,Mcdon64,Fisher06}, the proton analyzing
power $A_y$~\cite{All93,Vea01,Fisher06}, and other polarization
observables~\cite{Dan10}. The calculations
performed with a variety of NN interactions have shown a
glaring discrepancy between theory and experiment for 
$A_y$~\cite{Fon99,Vea01,HH03,Fisher06,DF07}. This discrepancy is very  
similar to the well known ``$A_y$  Puzzle''  in $N-d$ scattering. This is a
fairly old problem, already reported about 20 years
ago~\cite{KH86,WGC88} in the case of $n-d$ and later confirmed also in
the $p-d$ case~\cite{Kie96}. For other $p-\het$ observables, as the
$\het$ analyzing power $A_{0y}$ and some spin correlation observables,
 discrepancies have been also observed. 
Recently~\cite{Dan10} at the Triangle University National
Laboratory (TUNL)  there has been a new set of accurate 
measurements (at $E_p=1.60$, $2.25$, $4$ and $5.54$ MeV) of various
spin correlation coefficients, which has allowed for a phase-shift analysis (PSA).

In this letter we report a study of the effect of 3N force models
in $p-\het$ elastic scattering in order to see whether their
inclusion allows to reduce the above mentioned discrepancies. Clearly, it is
important to specify which NN potential is used together with a
particular model of 3N interaction. The N2LO 3N force derived from
EFT has been used together with the NN
potential models constructed within the same approach, in particular
the next-to-next-to-next-to-leading order (N3LO) interaction derived
by Entem and Machleidt~\cite{EM03,ME12}. We have considered two cutoff
values, $\Lambda=500$ MeV and $\Lambda=600$ MeV, 
labeled respectively N3LO500 and N3LO600. Correspondingly, we have to
fix the two parameters $c_D$ and $c_E$ present in the N2LO 3N
force. Together with 
the N3LO500 interaction model we have considered two versions of the
N2LO 3N force; in the first one, label-led N2LO500*, 
$c_D$ and $c_E$ have been chosen so as
to reproduce the $A=3,4$ binding energies as in Ref.~\cite{N07}.
In the second one, labeled N2LO500, the two parameters have been 
fixed reproducing the 3N binding energy and the tritium GTME~\cite{Mea12}. These two
models have been used to explore the dependence of the results on 
$c_D$ and $c_E$.

With the N3LO600 NN interaction model, we have considered the 3N
N2LO force label-led N2LO600 with  $c_D$ and $c_E$  fixed
to reproduce the 3N binding energy and the tritium GTME~\cite{Mea12}.
In this way we can explore the dependence on $\Lambda$ of 
the 4N observables. The specific values of the parameters $c_D$ and $c_E$ are
summarized in Table~\ref{tab:par}.

\begin{table}[t]
\caption{\label{tab:par}
  NN+3N interaction models used in this work. In columns 2$-$4 the values of
  the cutoff parameter $\Lambda$ and the coefficients $c_D$ and $c_E$ entering
  the EFT force models are reported (the coefficients are
  adimensionals). In the last column we have reported the
  corresponding $\heq$  binding energy.}
\begin{center}
\begin{tabular}{lcccccc}
\hline 
Model & $\Lambda$ [MeV]  & $c_D$ & $c_E$ & $B(\heq)$ [MeV] \\
\hline
N3LO500/N2LO500* & $500$ & $\pham 1.0\phan $ & $-0.029$ & $28.36$ \\
N3LO500/N2LO500  & $500$ & $-0.12$ & $-0.196$ & $28.49$ \\
N3LO600/N2LO600  & $600$ & $-0.26$ & $-0.846$ & $28.64$ \\
AV18/IL7 & & & & $28.44$ \\
\hline
\end{tabular}
\end{center}
\end{table}

The Illinois 3N model has been used in conjunction with the Argonne
$v_{18}$ (AV18) NN potential~\cite{AV18}. Between the different
Illinois models, we have considered the most recent one, the 
so called Illinois-7 model (IL7)~\cite{il7}. In Table~\ref{tab:par} we
have also reported the corresponding $\heq$ binding energy, which
results rather close to the experimental value of $28.30$
MeV. Therefore, eventual 4N forces should be rather tiny and their
effect in $p-\het$ scattering at low energy can be safely neglected.

For this study we have focused our attention to the effect of the 3N
interaction. For this reason we have restricted the electromagnetic interaction between
the nucleons to just the point Coulomb interaction between the
protons. To be noticed that with the AV18 potential one should
include the full electromagnetic interaction, including
two-photon exchange, Darwin-Foldy term, vacuum polarization, and
magnetic moment interactions as discussed in Ref.~\cite{AV18}. The effect of these
additional terms for $N-d$ scattering was studied in Refs.~\cite{Kie04,Mea09}
and found to have a sizeable effect for some polarization
observables. Regarding the N3LO500 and N3LO600 NN interactions, one should include
only the effect of the two-photon exchange, Darwin-Foldy term, and vacuum 
polarization interactions in the ${}^1S_0$
partial wave~\cite{ME12,MSV13}. Again, we have disregarded them in
this work. The effect of these additional electromagnetic
interactions will be the subject of a forthcoming paper~\cite{newHH}.

In the energy range considered here ($E_p\le 6$ MeV), the various
$p-\het$ observables are dominated by $S$-wave and $P$-wave phase shifts
($D$-wave phase shifts give only a marginal contribution, and more
peripheral phase shifts are negligible). A comparison of a
selected set of calculated phase-shifts and mixing parameters with those
obtained by the recent PSA~\cite{Dan10} reveals that, using the interaction models
with only a NN potential, both $S$- and $P$-wave phase-shifts result 
to be at variance with the PSA. Including the 3N force, we observe a general 
improvement of the description of the $S$- and $P$-wave phase shifts and mixing
parameters. A detailed comparison between the calculated phase-shifts
and those obtained from the PSA has been reported in Ref.~\cite{Fukuoka}.

\begin{figure}
  \includegraphics[width=\columnwidth,clip,angle=0]{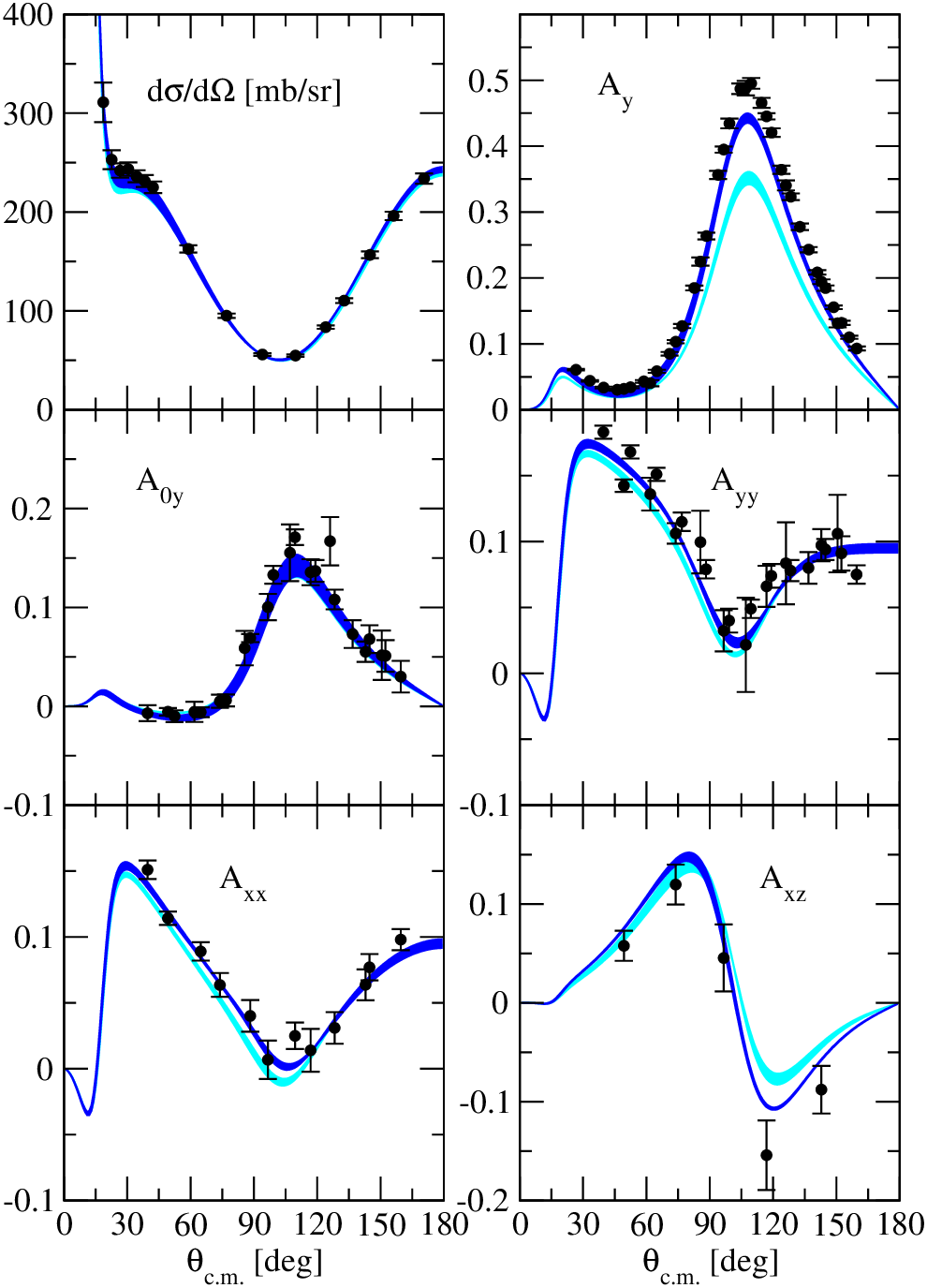}
  \caption{(color online) $p-\het$ differential cross section, analyzing powers and various
   spin correlation coefficients at $E_p=5.54$ MeV
    calculated with only the NN potential
    (light cyan band) or including also the 3N interaction (darker
    blue band). The experimental data are from
    Refs.~\protect\cite{All93,Vea01,Fisher06}.} 
   \label{fig:6obs}
\end{figure}

Let us compare the theoretical results directly with a
selected set of available experimental data.
To see the effect of the 3N interaction, we have reported in
Fig.~\ref{fig:6obs}  two bands, one
collecting the results obtained using only NN interaction models and
one obtained including also a 3N interaction.
We have reported the results for the $p-\het$ unpolarized differential cross section, two 
analyzing power observables, and some spin correlation observables.
We note that the differential cross
section, the $\het$ analyzing power $A_{y0}$, and the spin correlation
coefficients are not particularly sensitive to the adopted interaction models, 
and in general we observe a good agreement with the experimental values 
in all considered cases.

On the contrary, for the proton analyzing power $A_{y}$, shown in the upper right panel,
we note a large sensitivity to the inclusion of the 3N interaction.  
The calculations performed using N3LO500 and AV18, in fact, largely underpredict 
the experimental points, a fact already observed before~\cite{Vea01,Fisher06,Vea11}. A
sizable improvement is found by including the 3N interaction. 
The underprediction of the experimental data is now around 8-10\%. 

\begin{figure}
  \includegraphics[width=\columnwidth,clip]{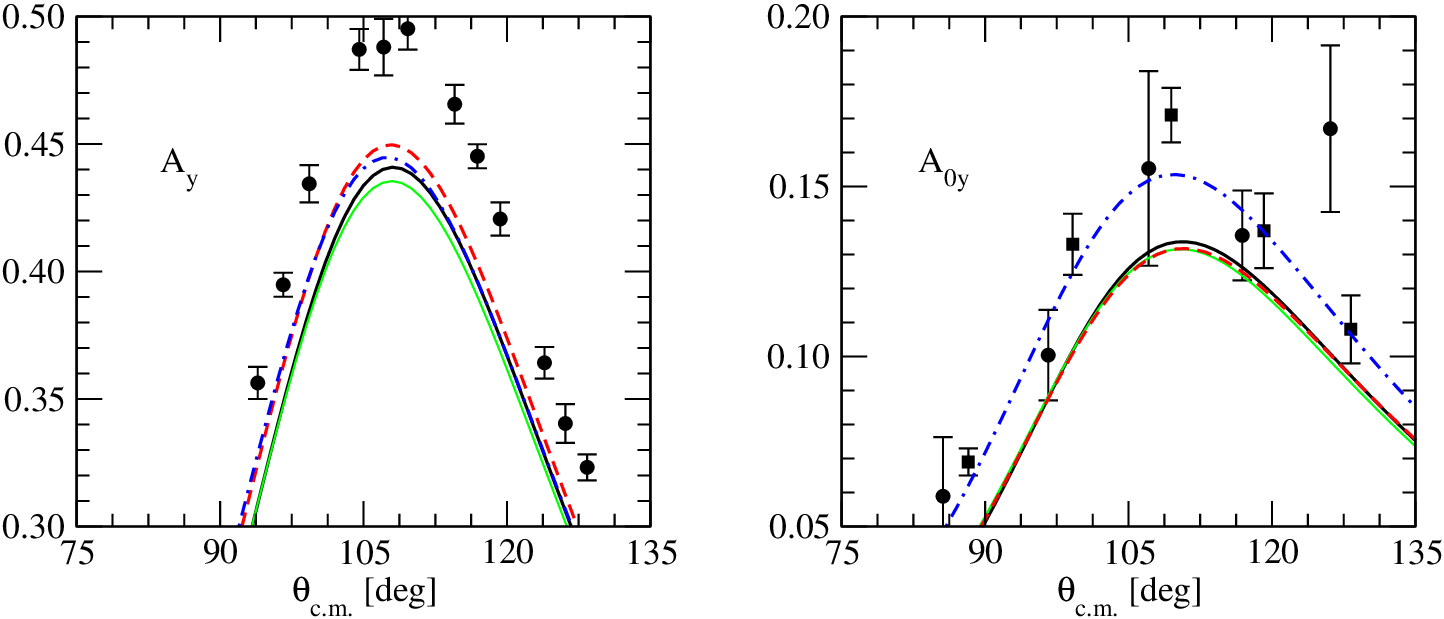}
  \caption{(color online) $p-\het$ observables at $E_p=5.54$ MeV 
   calculated with the N3LO500/N2LO500* (thick
    black solid lines), N3LO500/N2LO500 (thin green solid lines), N3LO600/N2LO600
   (dashed red lines), and AV18/IL7 (dot-dash blue lines) interaction models. 
   The experimental data are from
    Refs.~\protect\cite{All93,Vea01,Fisher06}.}
\label{fig:det}
\end{figure}

To better point out the sensitivity to the particular interaction model,
in Fig.~\ref{fig:det} an enlargement of $A_y$ and $A_{0y}$
in the peak region is shown. From the inspection of the
figure, we can see that the results obtained using the N3LO500/N2LO500*
and N3LO500/N2LO500 interaction models are very similar, showing that
there is not much sensitivity to the parameters $c_D$ and $c_E$. The
observables are more sensitive to the 
choice of the cutoff $\Lambda$, in particular $A_{y}$ calculated with 
the $\Lambda=600$ MeV interaction model is slightly closer to the 
experimental data. Finally, the $A_y$ calculated
with AV18/IL7 is very similar to those obtained with the chiral
models, while $A_{0y}$ is in better agreement with the 
data (however, for this observable the experimental uncertainties are
rather large).

The previously observed large underprediction of the $p-\het$ $A_y$
observable was considered to be due to some deficiencies of the
interaction in $P$-waves~\cite{Fon99,Vea01}, as, for example, 
due to the appearance of a unconventional ``spin-orbit'' interaction
in $A>2$ systems~\cite{K99}. The IL7 model has been fitted to
reproduce the $P$-shell nuclei spectra and, in particular, the two 
low-lying states in ${}^7$Li. This may explain the improvement in the
description of the $p-\het$ $A_y$ obtained with this interaction model.
Regarding the N2LO 3N force models, its two parameters have been
fitted either to the $A=3$ and $4$ binding energies, or to reproduce
the 3N binding energy and the tritium GTME, quantities which are more
sensitive to $S$-waves. Therefore, its capability to improve the
description of the $p-\het$ $A_y$ observable is not imposed but it is
somewhat built-in.  

\begin{figure}[tb]
 \includegraphics[clip,width=\columnwidth]{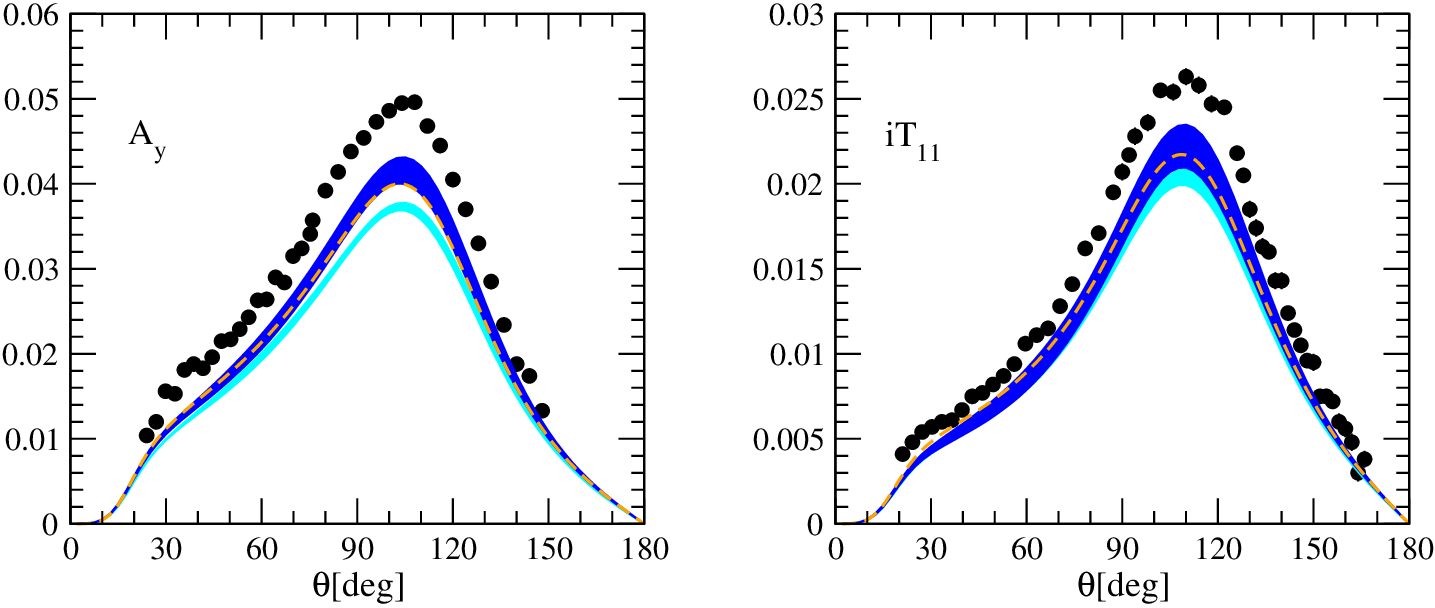}
 \caption{(color online) $p-d$ vector polarization observables at $E_p=3 $ MeV 
    calculated with only the NN potentials (light cyan band) or
    including also the 3N interactions (dark blue band) obtained
    within EFT. The results
    obtained with the AV18/IL7 interaction models are reported as the
    dashed (orange) lines. The experimental data
    are from Refs.~\protect\cite{Shimi95}. } 
\label{fig:pd}
\end{figure}

It is interesting to examine the effect of the same interaction models
in $p-d$ scattering. To this aim, we report in Fig.~\ref{fig:pd} two
vector polarization observables at $E_p=3$ MeV. In this figure, the
light (cyan) band has been obtained using the NN chiral interaction
only (in this case, the N3LO500 and N3LO600 models). The dark (blue)
band has been obtained adding the corresponding N2LO 3N interaction.
In this figure, the results obtained with AV18/IL7 are shown by the
dashed (orange) lines (in this case, we have included the effect of the
magnetic moment interactions since here it is sizable~\cite{Kie04}). As
can be seen, with the inclusion of 3N forces, the underprediction of both
observables is reduced, however it is still of the
order of 18-20\%, somewhat larger than for the $p-\het$ $A_y$
observable.  It should be noticed that the two $p-d$ asymmetries, though
rather tiny, show a large sensitivity to the $P$-waves phase-shift
splitting~\cite{Gea96,Kie96}. Accordingly, they can be used to fine tune the
strength of subleading 3N spin-orbit appearing at 
next-to-next-to-next-to-next-to-leading order (N4LO)~\cite{GKV11}.

In conclusion, we have presented for the first time an
analysis of $p-\het$ elastic scattering observables
including the effect of different 3N force models.
The results obtained have been compared with the available
experimental data. We have found that the phase shifts obtained with
both the chiral and AV18/IL7 models are very close~\cite{Fukuoka}
with those derived from the recent PSA performed at TUNL~\cite{Dan10}. The direct comparison of the
theoretical results with the experimental data has shown that there
are still some discrepancies, but the $A_y$ problem is noticeably
reduced. In fact, we observe that now the discrepancy is reduced to be
of the order of 10\% at the peak, much less than before. 
We have also found that the results obtained with the N3LO/N2LO models
and AV18/IL7 model are always rather close with each other (except for
$A_{0y}$). Since the
frameworks used to derive these 3N force models are rather
different, this outcome is somewhat surprising. 
Finally, it will be certainly very interesting to test the effect of 
the inclusion of the N3LO and N4LO 3N forces derived from EFT. Work in
this direction is in progress.

\begin{acknowledgments}
The Authors would like to acknowledge the assistance and help of the staff of the computer center of 
INFN-Pisa, where all the calculations presented in this paper were performed.
\end{acknowledgments}


\begin{thebibliography}{99}

\bibitem{Gea96}  W. Gl\"ockle {\it et al.},
Phys. Rep. {\bf 274}, 107 (1996)
%
\bibitem{CS98} J. Carlson and R. Schiavilla, Rev. Mod. Phys. {\bf 70},
    743 (1998)
%
\bibitem{KEMN12} N. Kalantar-Nayestanaki {\it et al.},
                 Rep. Prog. Phys. {\bf 75}, 016301 (2012)
%
\bibitem{Kolck94} U. van Kolck, Phys. Rev. C {\bf 49}, 2932 (1994)
%
\bibitem{Eea02} E. Epelbaum  {\it et al.},  Phys. Rev. C {\bf 66},
                064001  (2002) 
%
\bibitem{N07} P. Navr{\'a}til,  Few-Body Syst. {\bf 41}, 117 (2007) 
%
\bibitem{Gazit} A. Gardestig and D. R. Phillips, Phys. Rev. Lett. {\bf
  96}, 232301 (2006); D. Gazit, S. Quaglioni, and P. Navr\'atil,
   {\it ibid.} {\bf 103}, 102502 (2009).
%
\bibitem{Mea12} L.E. Marcucci {\it et al.},
               Phys.\ Rev.\ Lett. {\bf 108}, 052502 (2012)
%
\bibitem{ME12} R.\ Machleidt and D.R.\ Entem,
               Phys. Rep. {\bf 503}, 1 (2011)
%
\bibitem{BEKM11} V. Bernard V.  {\it et al.},
                 Phys. Rev. C {\bf 84}, 054001 (2011); 
                 {\it ibid.} {\bf 77},  064004 (2008)
%
\bibitem{KGE12} H. Krebs, A. Gasparyan, and E. Epelbaum,
                Phys. Rev. C {\bf 85}, 054006 (2012); 
                 {\it ibid.} {\bf 87},   054007 (2013)
%
\bibitem{GKV11} L. Girlanda, A. Kievsky, and M. Viviani,
                Phys. Rev. C {\bf 84}, 014001 (2011)
%
\bibitem{PPWC01} S. C. Pieper {\it et al.}, 
                 Phys. Rev. C {\bf 64}, 014001 (2001)
%
\bibitem{DF07} A. Deltuva and A. C. Fonseca, Phys. Rev. C {\bf 75},
               014005  (2007) 
%
\bibitem{DF07b} A. Deltuva and A. C. Fonseca, Phys. Rev. Lett. {\bf 98},
                162502  (2007); Phys. Rev. C {\bf 76},   021001    (2007)
%
\bibitem{Alt78} E. O. Alt, W. Sandhas, and H. Ziegelmann, Phys. Rev. C {\bf
    17},   1981 (1978); {\it ibid.} {\bf 21}, 1733 (1980) 
%
\bibitem{DFS05} A. Deltuva, A. C. Fonseca, and P.U. Sauer, Phys. Rev. C {\bf 71},
  054005 (2005); {\it ibid.} {\bf 72}, 054004 (2005) 
%
\bibitem{Cie98} F. Cieselski and J. Carbonell, Phys. Rev. C {\bf 58},
  58 (1998); F. Cieselski, J. Carbonell, and C. Gignoux, Phys. Lett. {\bf
    B447}, 199 (1999)
%
\bibitem{Lea05} R. Lazauskas  {\it et al.}, Phys. Rev. C {\bf 71},
                034004  (2005)  
%
\bibitem{HH03} H. M. Hofmann and G. M. Hale, Phys. Rev. C {\bf 68},
  021002 (2003); Phys. Rev. C {\bf 77}, 044002  (2008) 
%
\bibitem{Sofia08}  S. Quaglioni and P. Navr\'atil, Phys. Rev. Lett. {\bf 101},
  092501  (2008) 
%
\bibitem{rep08} A. Kievsky {\it et al.}, J. Phys. G:
   Nucl. Part. Phys. {\bf 35}, 063101  (2008) 
%
\bibitem{Vea11}  M. Viviani {\it et al.},
                 Phys. Rev. C {\bf 84}, 054010  (2011)
%
\bibitem{Fam54} K. F. Famularo {\it et al.}, Phys. Rev. {\bf 93},
  928  (1954) 
%
\bibitem{Mcdon64} D. G. McDonald, W. Haberli, and L. W. Morrow,
  Phys. Rev. {\bf 133}, B1178  (1964) 
%
\bibitem{Fisher06} B. M. Fisher {\it et al.}, Phys. Rev. C {\bf 74},
  034001  (2006) 
%
\bibitem{All93} M. T. Alley and L. D. Knutson, Phys. Rev. C {\bf 48},
  1890  (1993) 
%
\bibitem{Vea01} M. Viviani {\it et al.}, 
                Phys. Rev. Lett. {\bf 86}, 3739  (2001) 
%
\bibitem{Dan10} T.V. Daniels {\it et al.},
                Phys. Rev. C {\bf 82}, 034002 (2010)
%
\bibitem{Fon99} A. C. Fonseca, Phys. Rev. Lett. {\bf 83}, 4021  (1999) 
%
\bibitem{KH86} Y. Koike and J. Haidenbauer, Nucl. Phys. {\bf A463},
               365c  (1987) 
%
\bibitem{WGC88} H. Witala, W. Gl\"ockle, and T. Cornelius,
                Nucl. Phys. {\bf A491}, 157  (1988) 
%
\bibitem{Kie96} A. Kievsky  {\it et al.},
                 Nucl. Phys. {\bf A607}, 402  (1996); 
                A. Kievsky, M. Viviani, and S. Rosati, Phys. Rev. C
                 {\bf 64}, 024002 (2001)
%
\bibitem{EM03} D.R.\ Entem and R.\ Machleidt,
               Phys.\ Rev.\ C {\bf 68}, 041001 (2003) 
%
\bibitem{AV18} R.B.\ Wiringa, V.G.J.\ Stoks, and R.\ Schiavilla,
               Phys.\ Rev.\ C {\bf 51}, 38 (1995) 
%
\bibitem{il7} S. C. Pieper,
              AIP Conf. Proc. {\bf 1011}, 143 (2008)
%
\bibitem{Kie04} A. Kievsky, M. Viviani, and L. E. Marcucci
                Phys. Rev. C {\bf 69}, 014002 (2004)
%
\bibitem{Mea09} L. E. Marcucci  {\it et al.},
                Phys. Rev. C {\bf 80}, 034003 (2009)
%
\bibitem{MSV13} L.E. Marcucci, R. Schiavilla, and M. Viviani,
                Phys. Rev. Lett. {\bf 110}, 192503 (2013)
%
\bibitem{newHH} M. Viviani {\it et al.}: in preparation
%
\bibitem{Fukuoka}  M. Viviani {\it et al.}, {\tt  arXiv:1210.5890}
%
\bibitem{K99} A. Kievsky,
              Phys. Rev. C {\bf 60}, 034001 (1999)
%
\bibitem{Shimi95} S. Shimizu {\it et al.}, Phys. Rev. C {\bf 52}, 1193 (1995) 
%
\end{thebibliography}
\end{document}